\documentclass[aps,prd,reprint,superscriptaddress,showpacs,amsmath,amssymb]{revtex4-2}

\usepackage[colorlinks,
           bookmarks=true,
           linkcolor=blue,
           urlcolor=blue, 
            anchorcolor=black,
            citecolor=blue
            ]{hyperref}

\usepackage{multirow}
\usepackage{graphicx}
\usepackage{dcolumn}
\usepackage{bm}
\usepackage{float}
\usepackage{exscale}
\usepackage{relsize}
\usepackage{amsmath}
\usepackage{amssymb}
\begin{document}

\title{Macroscopic study of $\phi$-mesic nuclei in $pp$ collisions at 
$\sqrt{s}$ = 7 TeV with the PACIAE model}

\author{Qiang Wang}
\affiliation{Key Laboratory of Quark and Lepton Physics (MOE) and Institute of
	Particle Physics, Central China Normal University, Wuhan 430079,
	China}

\author{Zhi-Lei She}
\email[]{shezhilei@wtu.edu.cn}
\affiliation{School of Mathematics and Statistics, Wuhan Textile University, Wuhan 430200, China}

\author{An-Ke Lei}
\email{ankelei@gznu.edu.cn}
\affiliation{School of Physics and Electronic Science, Guizhou Normal University, Guiyang 550025, China}

\author{Dai-Mei Zhou}
\email[]{zhoudm@mail.ccnu.edu.cn, corresponding author}
\affiliation{Key Laboratory of Quark and Lepton Physics (MOE) and Institute of
	Particle Physics, Central China Normal University, Wuhan 430079,
	China}

\author{Wen-Chao Zhang}
\affiliation{School of Physics and Information Technology, Shaanxi Normal University, Xi'an 710119, China}

\author{Hua Zheng}
\affiliation{School of Physics and Information Technology, Shaanxi Normal University, Xi'an 710119, China}

\author{Li-Lin Zhu}
\affiliation{College of Physics, Sichuan University, Chengdu 610064, China}

\author{Yu-Liang Yan}
\affiliation{China Institute of Atomic Energy, P. O. Box 275 (10), Beijing 102413, China}

\author{Zhong-Qi Wang}
\affiliation{Nanjing University of Aeronautics and Astronautics, Nanjing 210000, China}

\author{Ben-Hao Sa}
\email[]{sabhliuym35@qq.com}
\affiliation{China Institute of Atomic Energy, P. O. Box 275 (10), Beijing 102413, China}
\affiliation{Key Laboratory of Quark and Lepton Physics (MOE) and Institute of Particle Physics, Central China Normal University, Wuhan 430079, China}
\date{\today}

\begin{abstract}
A macroscopic method of the parton and hadron cascade model PACIAE combined with the 
coalescence model DCPC is employed to study the $\phi$-mesic nuclei in $pp$ 
collisions at $\sqrt{s}$ = 7 TeV for the first time in this work. Based on 
the final hadronic states of $pp$ collisions simulated by the PACIAE model, the 
$\phi$-mesic nuclei: $\mathrm{^3_{\phi} H}(\phi-pn)$, 
$\mathrm{^4_{\phi} H}(\phi-pnn)$, and $\mathrm{^4_{\phi} He}(\phi-ppn)$ are recombined 
by the DCPC model with constituents of protons, neutrons, and $\phi$ mesons. 
Their yields, transverse momentum spectra, and rapidity distributions are 
simulated separately. These theoretical predictions call for 
experimental confirmation.
\end{abstract}

\maketitle

\section{Introduction}

The interaction between $\phi$ mesons and nucleons is one of the most 
interesting and important issues in nuclear physics \cite{Hirenzaki:2010zza,Yamagata-Sekihara:2010ell,Cobos-Martinez:2017woo,Gao:2017hya} because the 
$\phi$-mesic nuclei are exotic few-body systems with strong 
interaction \cite{Krein:2017usp,Metag:2017yuh}. Due to the OZI 
rule \cite{Okubo:1963fa,Zweig:1964jf,Iizuka:1966fk}, the direct coupling 
between a $\phi$ meson ($s\bar{s}$) and a nucleon is expected to be suppressed. The interaction between a $\phi$ meson and a nucleon can not be fully explained by 
the light-meson exchange theory. Its explanation requires other mechanisms such as the QCD van der 
Waals interaction ~\cite{Appelquist:1978rt,Ellis:1988jwa}. Therefore, the 
$\phi$-mesic nuclei provide a unique laboratory for studying the properties 
of the interaction between $\phi$ meson and nucleon and provide insight into the theory of 
the strong interaction in the non-perturbative regime. 

Recently, the ALICE Collaboration has measured the $p-\phi$ momentum 
correlation function and indicated the attractive $p-\phi$ interaction which 
results in the possible $p-\phi$ bound state~\cite{ALICE:2021cpv}. The $N-\phi$ potential 
of the $^4 S_{3/2}$ channel based on the (2+1)-flavor lattice QCD calculations has 
been simulated by the HAL QCD Collaboration \cite{Lyu:2022imf}. Subsequently, 
the $p-\phi$ interaction of the $^4 S_{3/2}$ channel was used to constrain the 
$^2 S_{1/2}$ channel indicating the possible bound state of the $N-\phi(^2 S_{1/2})$ 
system~\cite{Chizzali:2022pjd}.

More recently, these studies have been extended to $\phi$-mesic few-body 
systems~\cite{Belyaev:2007yc,Etminan:2024vkv,Filikhin:2024avj,Belyaev:2009ag,Sofianos_2010,Wen:2025wit,Lazauskas:2026ysy}. The microscopic approach based on the 
Faddeev–Yakubovsky equations has been used to calculate binding energies of 
$\phi$-mesic nuclei, predicting the possible existence of $\mathrm{^4_{\phi} H}(\phi-pnn)$, 
$\mathrm{^4_{\phi} He}(\phi-ppn)$, and 
$\mathrm{^5_{\phi} He}(\phi-ppnn)$~\cite{Lazauskas:2026ysy}.

In contrast to the microscopic studies of $\phi$-mesic nuclei mentioned above, we use a 
macroscopic method of the parton and hadron cascade model PACIAE combined with the 
dynamically constrained phase-space coalescence DCPC model to investigate 
the production of $\phi$-mesic nuclei in $pp$ collisions at $\sqrt{s}$ = 7 TeV 
for the first time. To this end, the PACIAE 4.0 model~\cite{Lei:2024kam} is 
first used to simulate the final hadronic state (FHS) of $pp$ collisions. The 
$\mathrm{^3_{\phi} H}(\phi-pn)$, $\mathrm{^4_{\phi} H}(\phi-pnn)$, and 
$\mathrm{^4_{\phi} He}(\phi-ppn)$ are then recombined with constituent particles 
of protons, neutrons, and $\phi$ mesons in the above FHS by the DCPC 
model~\cite{Yan:2011fq}. Furthermore, the yields, transverse momentum 
($p_T$) spectra and rapidity ($y$) distributions are calculated for 
$\mathrm{^3_{\phi} H}$, $\mathrm{^4_{\phi} H}$, and $\mathrm{^4_{\phi} He}$, separately. 

\section{The PACIAE model and related DCPC model} 

In the parton and hadron cascade model PACIAE 4.0 \cite{Lei:2024kam}, 
PYTHIA 8.3 \cite{Bierlich:2022pfr} is first used to simulate the 
nucleon-nucleon (NN) and/or hadron-hadron (hh) collisions with the 
string fragmentation temporarily switched off. The initial partonic state 
composed of quarks, antiquarks, and gluons is then produced. It is 
followed by the partonic rescattering, which results in a final partonic state 
(FPS) filled with abundant quarks, antiquarks and gluons. For the subsequent 
hadronization, there are Lund string fragmentation mechanism and the 
phenomenological coalescence model \cite{Lei:2024kam} available for selection. 
The former is chosen in this work. After $2 \to 2$ hadronic 
rescattering~\cite{Lei:2024kam} is performed, the final hadronic state (FHS) 
composed of numerous hadrons with their four-coordinate and four-momentum is 
produced. 

In the statistical mechanics \cite{Thermo_Cambri,Thermo_North}, the yield of an
$N$-particle cluster is defined as
\begin{equation}
Y_N = \int ... \int_{E_a \le H \le E_b } \frac{ \mathrm{d}\vec{r}_1 \mathrm{d}\vec{p}_1 ...
		\mathrm{d}\vec{r}_N \mathrm{d}\vec{p}_N }{h^{3N} } ,
\end{equation}
where $E_a$ and $E_b$ are, respectively, the lower and upper energy thresholds 
of the cluster. $\vec{r}_i$ and $\vec{p}_i$ denote, respectively, 
the three-coordinate and three-momentum of the $i$-th particle in the 
center-of-mass system to which the cluster belongs. 

Therefore, the yield of $\mathrm{^3_{\phi} H}$ (cluster), for instance, can be 
expressed in terms of its components of $\phi$, $p$, and $n$ 
(particles) as
\begin{equation}
Y_{\mathrm{^3_{\phi} H}} = \int ... \int \delta_{123} \prod_{i=1}^{3} 
	               \frac{\mathrm{d}\vec{r}_i \mathrm{d}\vec{p}_i } {h^3},
\end{equation}
where
\begin{equation}
	\delta_{123} =
	\begin{cases}
		1  \quad \mathrm{if} \quad 1 \equiv \phi, \quad 2 \equiv p, \quad 3 \equiv n \\
		\qquad m_0-\Delta m \le m_{\mathrm{inv} } \le m_0+\Delta m,  \\
		\qquad r_{i0} \le R_0, (i=1,2,3)  \\
		0  \quad \mathrm{otherwise}.
	\end{cases}\label{eq:DCPC_constr_1_ENG}
\end{equation}
Here the $\delta_{123}$ function is introduced in order for the cluster 
($\mathrm{^3_{\phi} H}$) to exist naturally but not statistically. The $m_0$ 
and $\Delta m$ refer to the mass of $\mathrm{^3_{\phi}H}$ and its mass 
uncertainty. They are assumed to be, respectively, the sum of the masses of its component hadrons~\cite{ParticleDataGroup:2024cfk} and the binding energies 
of $\mathrm{^3_{\phi}H}$~\cite{Lazauskas:2026ysy} as given in 
Table~\ref{table_m0_Delta_m}. The $R_0$ and $r_{i0}$ denote, respectively, the 
free radius parameter of $\mathrm{^3_{\phi}H}$ and the relative distance 
between the $i$-th component hadron and the center of mass of the 
$\mathrm{^3_{\phi}H}$ cluster. We set three ranges of [1.0, 2.0], [1.0, 3.0], 
and [1.0, 5.0] fm for $R_0$. Here the lower bound of 1.0 fm is the 
typical hadronic size \cite{Schweitzer:2012hh,Geiger:1994iq,Ellis:1995pe}. 	
The invariant mass $m_{\mathrm{inv}}$ is expressed as
\begin{equation}
m_{\mathrm{inv} } = \sqrt{\left(\sum_{i=1}^{3} E_i\right)^2-
	        \left( \sum_{i=1}^{3} \vec{p}_i \right)^2},
\end{equation}
where $E_i$ and $\vec{p}_i$ denote, respectively, the energy and 
three-momentum of the $i$-th component hadron.

In order to perform a macroscopic study of the production of the $\phi$-mesic 
nuclei in $pp$ collisions at $\sqrt{s}$ = 7 TeV, we first simulate a large 
number of final hadronic states for $pp$ collisions using the PACIAE model. 
Based on these FHSs, the component hadron list composed of $\phi$, $p$, and 
$n$ is then generated. Lastly we construct a four-layer loop over the 
component hadron list for the generation of $\mathrm{^4_{\phi} H}$ and 
$\mathrm{^4_{\phi} He}$, and a three-layer loop over component hadron list for 
the production of $\mathrm{^3_{\phi} H}$.  

\begin{table}[ ]
\caption{The values of the mass $m_0$ and its uncertainty parameter $\Delta m$ 
         for $\mathrm{^3_{\phi} H}$, $\mathrm{^4_{\phi} H}$, and 
	 $\mathrm{^4_{\phi} He}$.}
	\label{table_m0_Delta_m}
	\centering
	\setlength{\tabcolsep}{8pt}
	\renewcommand{\arraystretch}{1.2}
	\begin{tabular}{ccc}
		\hline
		\hline
		$\phi$-mesic nuclei & $m_0$ (MeV/$c^2$) & $\Delta m$ (MeV/$c^2$) \\
		\hline
		$\mathrm{^3_{\phi} H}$ & 2897  & 50.26  \\
		$\mathrm{^4_{\phi} H}$ & 3837 & 55.55  \\
		$\mathrm{^4_{\phi} He}$ & 3836  & 54.71  \\

		\hline
		\hline						
	\end{tabular}				
\end{table}

\section{Results and discussion} 

\begin{figure}[h]
\centering
\includegraphics[width=0.50\textwidth]{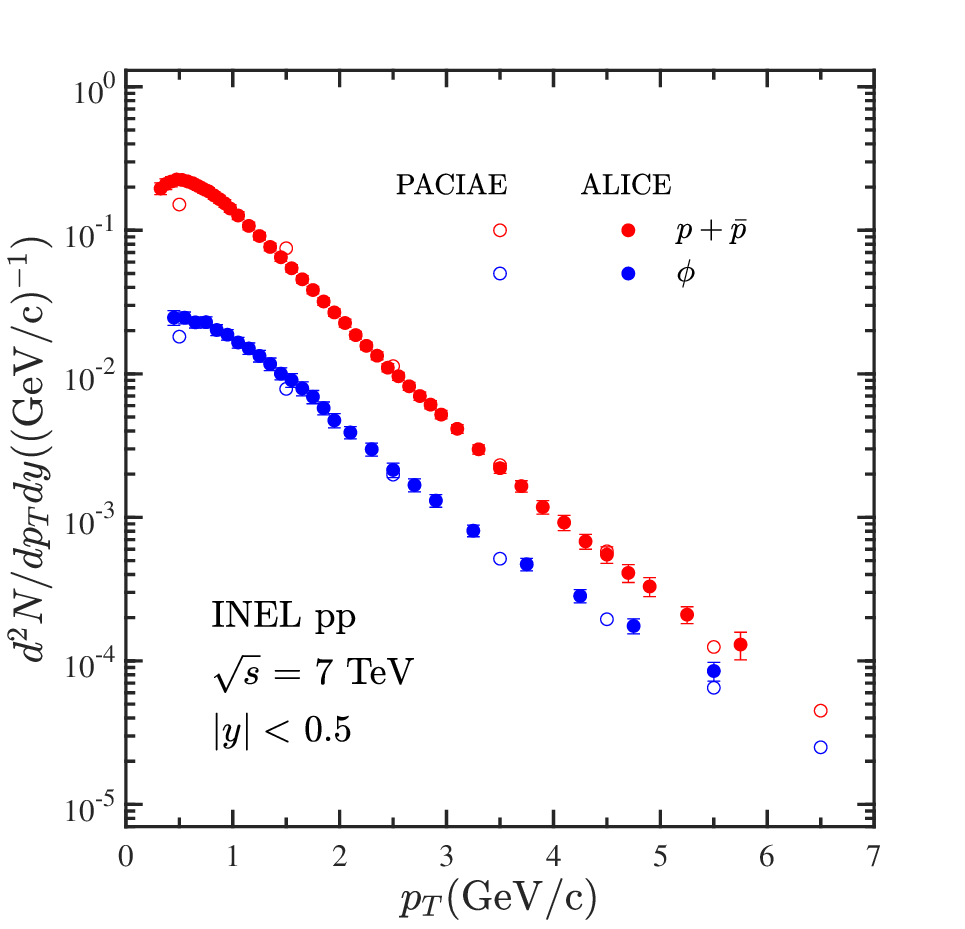}
\caption{ \centering{The fitted $p_T$ spectra of $p+\bar{p}$ and $\phi$  
(empty markers) and the corresponding ALICE data (solid 
markers)~\cite{ALICE:2015ial,ALICE:2012pjb} in $pp$ collisions at 
$\sqrt{s}$ = 7 TeV.} }
\label{fig_phi_N_dN_dpT}
\end{figure}

\begin{figure}[h]
\centering
\includegraphics[width=0.50\textwidth]{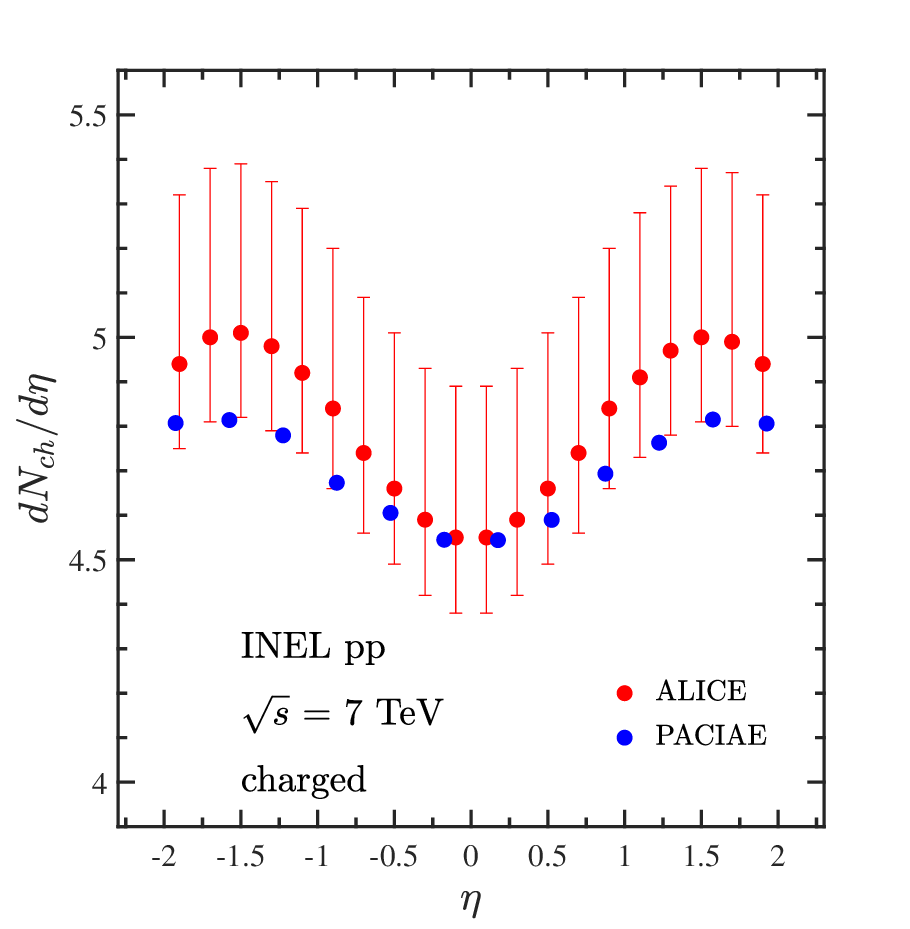}
\caption{ \centering{The fitted pseudorapidity distributions of charged 
particles (blue markers) and the corresponding ALICE data~\cite{ALICE:2015olq}
(red markers) in $pp$ collisions at $\sqrt{s}$ = 7 TeV. }
	}\label{fig_ch_dN_deta_errorbar}
\end{figure}

The PACIAE 4.0 model is used to generate 600 million final hadronic states for 
$pp$ collisions at $\sqrt{s}$ = 7 TeV. The three main parameters in the PACIAE 
model, adj1(10) (the $K$ factor multiplying the differential cross sections 
for hard parton-parton process), adj1(34) (the width of $p_T$ sampling) and 
probStoUD (the parameter for the suppression of selecting an $s\bar{s}$ pair 
with respect to $u\bar{u}$ and $d\bar{d}$) are adjusted to be adj1(10) = 0.8, 
adj1(34) = 0.46 and probStoUD = 0.250 by fitting the $p_T$ spectra of 
$p+\bar{p}$ and $\phi$ measured by ALICE 
Collaboration \cite{ALICE:2015ial,ALICE:2012pjb}. Other parameters in the 
PACIAE model are the default values. The fitted $p_T$ spectra along 
with the corresponding experimental data are shown in 
Fig.~\ref{fig_phi_N_dN_dpT}. Fig. \ref{fig_ch_dN_deta_errorbar} 
gives the fitted pseudorapidity distributions of charged particles and the 
corresponding ALICE data~\cite{ALICE:2015olq}. It can be seen that the PACIAE model  
results reproduce the ALICE pseudorapidity distributions within the error bars.

\begin{table}[ ]
	\caption{The simulated yields of $\mathrm{^3_{\phi} H}$, 
	$\mathrm{^4_{\phi} H}$, and $\mathrm{^4_{\phi} He}$ for three $R_0$ 
	ranges of [1.0, 2.0], [1.0, 3.0] and [1.0, 5.0] fm.}
	\label{table_yield_R0}
	\centering
	\setlength{\tabcolsep}{5pt}
	\renewcommand{\arraystretch}{1.2}
	\begin{tabular}{cccc}
		\hline
		\hline
		$\phi$-mesic nuclei & [1.0, 2.0] fm & [1.0, 3.0] fm & [1.0, 5.0] fm \\
		\hline
		$\mathrm{^3_{\phi} H}$ & $1.92 \times 10^{-6}$   & $9.50 \times 10^{-6}$ &  $2.66 \times 10^{-5}$\\
		$\mathrm{^4_{\phi} H}$ & $2.02 \times 10^{-8}$   & $2.09 \times 10^{-7}$ &  $9.25 \times 10^{-7}$\\
		$\mathrm{^4_{\phi} He}$ & $2.44 \times 10^{-8}$   & $2.06 \times 10^{-7}$ &  $8.36 \times 10^{-7}$\\		
		
		\hline
		\hline						
	\end{tabular}				
\end{table}

\begin{figure*}[ ]
\centering
\includegraphics[width=1.00\textwidth]{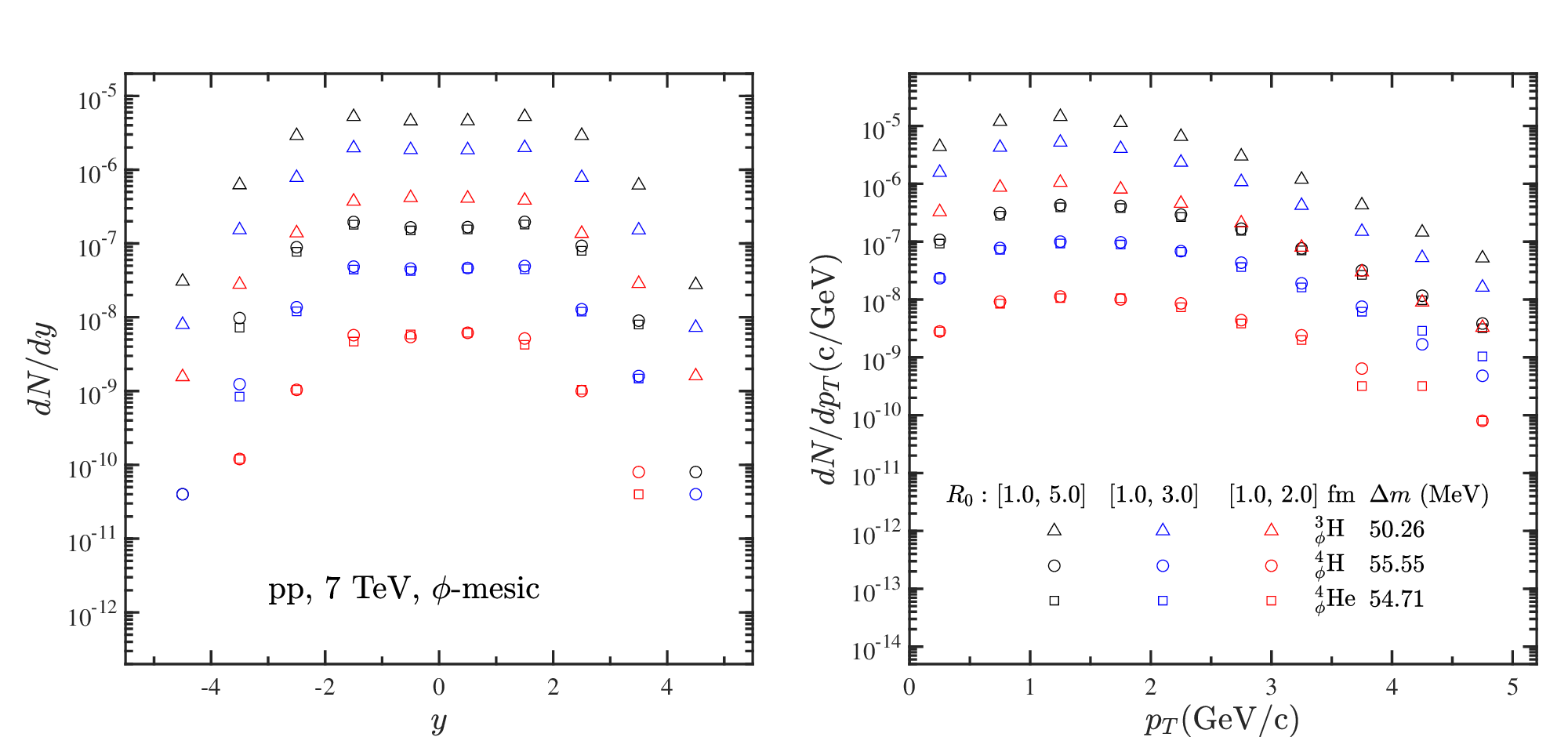}
\caption{ \centering{The simulated $y$ distributions (left panel) and $p_T$ spectra (right panel) 
of $\mathrm{^3_{\phi} H}$ (triangles), $\mathrm{^4_{\phi} H}$ (circles), and $\mathrm{^4_{\phi} He}$ (squares) in $pp$ 
collisions at $\sqrt{s}$ = 7 TeV. The black, blue, and red markers 
refer to the results simulated in the $R_0$ ranges of [1.0, 5.0], [1.0, 3.0], 
and [1.0, 2.0] fm, respectively.}}
\label{fig_phiN_display_pT_y}
\end{figure*}

The simulated yields of $\mathrm{^3_{\phi} H}$, $\mathrm{^4_{\phi} H}$, and 
$\mathrm{^4_{\phi} He}$ are shown in Table~\ref{table_yield_R0} for three$R_0$ ranges of [1.0, 2.0], [1.0, 3.0] and [1.0, 5.0] fm. The simulated $y$ 
distributions (left panel) and $p_T$ spectra (right panel) of 
$\mathrm{^3_{\phi} H}$, $\mathrm{^4_{\phi} H}$, and $\mathrm{^4_{\phi} He}$ 
in $pp$ collisions at $\sqrt{s}$ = 7 TeV are shown in 
Fig.~\ref{fig_phiN_display_pT_y}. 
The assumed values of $m_0$ and $\Delta m$ are given in 
Table~\ref{table_m0_Delta_m}.
In Fig.~\ref{fig_phiN_display_pT_y}, the black, blue, and red markers refer to  
the simulated results with assumed $R_0$ ranges of [1.0, 5.0], [1.0, 3.0] and 
[1.0, 2.0] fm, respectively. The triangles, circles, and squares 
refer to the $\mathrm{^3_{\phi} H}$, $\mathrm{^4_{\phi} H}$, and $\mathrm{^4_{\phi} He}$, 
respectively.

The above simulated results show that the yields of the two four-body systems 
($\mathrm{^4_{\phi} H}$ and $\mathrm{^4_{\phi} He}$) are much lower than that 
of the three-body systems ($\mathrm{^3_{\phi} H}$). But the yields of the two 
four-body systems ($\mathrm{^4_{\phi} H}$ and $\mathrm{^4_{\phi} He}$) are 
almost equal. $\mathrm{^4_{\phi} H}$ and $\mathrm{^4_{\phi} He}$ can be 
regarded as the composite systems composed of core nucleus of $\mathrm{^3 H}$ 
and $\mathrm{^3 He}$ with one $\phi$ meson, respectively. Since the isospin of 
the $\phi$ meson is equal to zero, the $\mathrm{^4_{\phi} H}$ and the 
$\mathrm{^4_{\phi} He}$ satisfy the isospin symmetry of the mirror nuclei. It 
is observed that the yields of $\phi$-mesic nuclei increase with the 
increasing of the $R_0$ range from [1.0,2.0], to [1.0, 3.0], and to 
[1.0,5.0] fm.

In summary, based on the final hadronic states simulated by the PACIAE model 
in $pp$ collisions at $\sqrt{s}$ = 7 TeV, we recombined the $\phi$-mesic 
nuclei: $\mathrm{^3_{\phi} H}$, $\mathrm{^4_{\phi} H}$, and 
$\mathrm{^4_{\phi} He}$ by the DCPC model. The simulated yields of  
$\mathrm{^4_{\phi} H}$ and $\mathrm{^4_{\phi} He}$ are much lower than that of
$\mathrm{^3_{\phi} H}$, while the yields of $\mathrm{^4_{\phi} H}$ and 
$\mathrm{^4_{\phi} He}$ are almost identical. As predictions, the simulated 
transverse momentum ($p_T$) spectra and rapidity ($y$) distributions of 
$\phi$-mesic nuclei require experimental verification. 

Further research on the $\phi$-mesic nuclei with five or more bodies is 
necessary. Extending to $p$-Pb and Pb-Pb collisions is also required. New 
observables such as elliptic flow and nuclear modification factor are 
also worthy of investigation.

\section*{acknowledgments}

This work is supported by the National Natural Science
Foundation of China under Grants   12375135. Y.L.Y. acknowledges the Continuous Basic
Scientific Research Project (Grant  WDJC-2019-13). The work of W.C.Z. is supported
by the Natural Science Basic Research Plan in Shaanxi Province of
China(Program  2026JC-YBMS-0078). 
A. K. L. acknowledges the project of Guizhou Provincial Department of Science and Technology under Grant  ZD[2026]108.


%

\end{document}